\documentclass[aps,prd,twocolumn,superscriptaddress,groupedaddress]{revtex4-2}

\usepackage{graphicx}
\usepackage{amsmath}
\usepackage{amssymb}
\usepackage{bm}
\usepackage{physics}
\usepackage{hyperref}

\begin{document}

\title{Operational measurement of relativistic equilibrium from stochastic fields alone}

\author{Ira Wolfson}
\email{wolfsoni@braude.ac.il}
\affiliation{Department of Electrical and Electronic Engineering,
Braude Academic College of Engineering, Karmiel, Israel}

\date{\today}

\begin{abstract}
The inverse-temperature four-vector $\beta^\mu = u^\mu/(k_B T_0)$
has been the theoretically accepted description of relativistic
equilibrium since the work of van~Kampen and Israel, yet no experiment
has ever reconstructed $\beta^\mu$ as a single observable.
All existing methods---Thomson scattering, spectral fitting,
blast-wave models---infer rest-frame temperature and flow velocity
from separate measurements and model-dependent assumptions.
We propose the first protocol that extracts both components
of $\beta^\mu$ from the \emph{same} passive observable:
electromagnetic fluctuation correlations emitted by a drifting medium.
A dimensionless $E$--$B$ cross-spectral ratio yields the drift velocity
directly from the Lorentz mixing of the field-strength tensor,
while angle-resolved noise power governed by the covariant
fluctuation--dissipation theorem provides the rest-frame temperature
through a ratio method that cancels absolute amplitude.
Together, these observables reconstruct $\beta^\mu$
without external probes, spectral lines, or absolute radiometric
calibration.
The protocol also enables the first direct experimental test of
whether the thermal state of a relativistic medium transforms
as a four-vector under Lorentz boosts---a question that has remained
purely theoretical since the Planck--Ott--Landsberg controversy
began in 1907.
Such a verified baseline is a prerequisite for assessing
temperature measurements in systems where $\beta^\mu$ cannot be
independently checked: in relativistic heavy-ion collisions,
the four-vector transformation is assumed but never tested;
in astrophysical sources such as gamma-ray bursts, the principle
of extracting thermal parameters from passive fluctuation
statistics could inform single-sightline analyses once the
multi-angle protocol has been validated.
Monte Carlo simulations parameterized to the HIGGINS dual 100\,TW
laser-plasma facility demonstrate sub-percent rest-frame temperature
recovery for Lorentz factors $\gamma = 1.05$--$10$,
with robustness to additive noise at signal-to-noise ratios $\gtrsim 10$.
The method requires relative channel calibration, knowledge of the
medium response function, and local thermodynamic equilibrium,
but eliminates the need for absolutely calibrated radiometers,
active probes, or spectral line identification.
\end{abstract}

\maketitle

\section{Introduction}
\label{sec:intro}

Relativistic thermodynamic equilibrium is described not by a scalar
temperature but by the inverse-temperature four-vector
$\beta^\mu = u^\mu/(k_B T_0)$~\cite{Becattini2016,Gavassino2022}.
This covariant structure, originating in the nonequilibrium statistical
operator formalism of Zubarev~\cite{Zubarev1979} and placed on rigorous
footing by van~Kampen~\cite{vanKampen1968} and
Israel~\cite{Israel1979}, now underpins relativistic dissipative
hydrodynamics and the Kubo--Martin--Schwinger condition in thermal
field theory~\cite{KapustaGale,LeBellac,Cercignani2002}.
Despite this theoretical consensus, $\beta^\mu$ has never been
reconstructed as a unified quantity from a single experimental
measurement.
All current techniques infer the rest-frame temperature $T_0$ and the
four-velocity $u^\mu$ from independent observables, and each
approach carries structural limitations that prevent
model-independent access to the covariant thermal state:

\begin{itemize}

\item \emph{Thomson scattering}~\cite{Sheffield2011} provides
high-accuracy electron temperature measurements but requires an
external laser probe and determines the flow velocity from an
independent Doppler shift measurement.
Temperature and velocity are separate observables, obtained from
different physical mechanisms, assembled into $\beta^\mu$ only
after the fact.

\item \emph{Spectral radiometry}~\cite{Hutchinson2002,RybickiLightman1979}
infers temperature by fitting thermal emission spectra, requiring
absolutely calibrated detectors and knowledge of the emission mechanism.
Drift velocity must be supplied independently,
typically from Doppler-shifted spectral lines.

\item \emph{Heavy-ion methods} (blast-wave
fits~\cite{BuszaRajagopal2018}, thermal dilepton
spectra) extract temperature from invariant-mass
distributions and velocity from collective flow observables ($v_1$,
$v_2$), but both require extensive modeling of freeze-out geometry,
equation of state, and non-equilibrium evolution.
The resulting $\beta^\mu$ is a fit parameter of a hydrodynamic
model, not a direct observable.

\item \emph{Astrophysical observations}~\cite{KumarZhang2015}
of relativistic jets and fireballs provide single-line-of-sight data
only, making the angle-resolved measurement required to reconstruct
a four-vector geometrically impossible.
Cosmological and gravitational redshifts are degenerate with thermal
Doppler shifts, preventing model-independent separation of $T_0$
and $\beta$.

\end{itemize}

A deeper consequence of this observational gap is that the
century-old question of how thermodynamic quantities transform
under Lorentz boosts has never been directly tested.
The Planck--Ott--Landsberg
controversy~\cite{Planck1908,Ott1963,Landsberg1966}---whether a
moving body appears hotter, cooler, or unchanged---was resolved
theoretically by recognizing that temperature is not a scalar but a
component of
$\beta^\mu$~\cite{vanKampen1968,Israel1979,Becattini2016}.
But this resolution rests entirely on theoretical arguments and
thermodynamic consistency; no experiment has ever measured the thermal
state of a relativistic medium at multiple observation angles and
verified that the results are consistent with four-vector
transformation.
(Angle-resolved synchrotron radiation from storage rings, while
relativistic and well-characterized, is coherent emission from
individual accelerating charges, not equilibrium thermal
fluctuations; it tests classical electrodynamics of single
particles, not statistical-mechanical properties of a thermal
ensemble.)
A protocol that reconstructs $\beta^\mu$ from angular measurements
would simultaneously serve as the first direct empirical test of this
covariant structure.

The implications extend beyond the laboratory. In relativistic
heavy-ion collisions, temperature and flow velocity are extracted
from particle spectra under the assumption that the thermal state
transforms as $\beta^\mu$---but this assumption is built into the
hydrodynamic analysis framework, not tested by it. A controlled
laboratory measurement that confirms (or falsifies) the four-vector
transformation law provides the empirical baseline against which
heavy-ion thermal analyses can be critically assessed. Similarly,
in astrophysical settings such as gamma-ray bursts, the principle
of thermal-state reconstruction from passive fluctuation statistics
could inform single-sightline spectral analyses, though adapting the
multi-angle protocol to single-line-of-sight data would require
further theoretical development.
The laboratory measurement is therefore not a proof of concept for
its own sake: it is the necessary first step in a chain that extends
from plasma diagnostics to high-energy nuclear physics and
relativistic astrophysics.

Passive thermal noise has been used to infer bulk drift velocity
in nonrelativistic space plasmas via quasi-thermal noise
spectroscopy~\cite{Issautier1999}, and full electromagnetic
fluctuation tensors for drifting relativistic plasmas have been
computed from kinetic theory~\cite{Ruyer2013}, but no prior work
has exploited $E$--$B$ cross-correlations or formulated an
operational reconstruction of $\beta^\mu$ from stochastic field
statistics.

We show that passive electromagnetic fluctuations emitted by a
drifting thermal medium encode both components of $\beta^\mu$ in their
correlation structure.
Two observables, extracted from the same stochastic field ensemble,
provide this information:
(i) the dimensionless cross-spectral ratio
$R = \langle \delta E'_y \delta B'^*_x \rangle / \langle |\delta E'_y|^2 \rangle$,
which yields the drift velocity from the Lorentz mixing of the
field-strength tensor without any intensity calibration; and
(ii) angle-resolved noise spectra governed by the covariant
fluctuation--dissipation theorem, from which the rest-frame temperature
is recovered via a ratio method that cancels absolute amplitude.
The $E$--$B$ cross-correlation is a genuinely new observable with no
counterpart in conventional radiometry, Stokes polarimetry, or
Thomson diagnostics.

The relationship to existing physics is analogous to that of
Johnson--Nyquist noise thermometry~\cite{Johnson1928,Nyquist1928} to
Planck's radiation law: the underlying thermal physics---Planck's
spectrum---was already understood, but the operational
protocol of measuring voltage noise across a resistor rather than
spectral radiance from a cavity constituted a distinct and practically
important diagnostic~\cite{CallenWelton1951}.
Here, the covariant fluctuation--dissipation theorem is well
established; the contribution is the identification of observables
and a complete measurement protocol that provides operational access
to $\beta^\mu$ for the first time.

We characterize the statistical precision and noise robustness of
this protocol using Monte Carlo simulations parameterized to the
HIGGINS dual 100\,TW laser-plasma facility at the Weizmann Institute
of Science~\cite{Kroupp2022}.

\section{Velocity extraction from fluctuation cross-correlations}
\label{sec:velocity}

We begin from the electromagnetic field-strength tensor
\begin{equation}
F^{\mu\rho} = \begin{pmatrix}
0 & -E_x & -E_y & -E_z \\
E_x & 0 & -B_z & B_y \\
E_y & B_z & 0 & -B_x \\
E_z & -B_y & B_x & 0
\end{pmatrix},
\label{eq:Fmunu}
\end{equation}
Under a Lorentz boost with velocity $v=\beta c$ along $\hat{z}$, the field tensor transforms as $F'^{\mu\rho} = \Lambda^\mu{}_\alpha\,\Lambda^\rho{}_\beta\,F^{\alpha\beta}$, where
\begin{equation}
\Lambda^\mu{}_\rho = \begin{pmatrix}
\gamma & 0 & 0 & -\gamma\beta \\
0 & 1 & 0 & 0 \\
0 & 0 & 1 & 0 \\
-\gamma\beta & 0 & 0 & \gamma
\end{pmatrix}.
\label{eq:boost}
\end{equation}
The transverse components of the boosted tensor mix electric and magnetic fields:
\begin{align}
F'^{02} &= \gamma(F^{02} + \beta\,F^{32}) = -\gamma(E_y + vB_x), \label{eq:F02}\\
F'^{13} &= \gamma(F^{13} + \beta\,F^{10}) = -\gamma\!\left(B_x + \frac{v}{c^2}E_y\right), \label{eq:F13}
\end{align}
which yield the standard field transformations
\begin{align}
E'_y &= \gamma(E_y + vB_x), \label{eq:boost_E}\\
B'_x &= \gamma\!\left(B_x + \frac{v}{c^2}E_y\right). \label{eq:boost_B}
\end{align}
Equations~(\ref{eq:boost_E})--(\ref{eq:boost_B}) are the key observation: the boost mixes the $(E_y, B_x)$ sector of $F^{\mu\rho}$, so that even if $E_y$ and $B_x$ are uncorrelated in the rest frame, their laboratory-frame counterparts $E'_y$ and $B'_x$ acquire a cross-correlation proportional to $v$.

In the rest frame of an isotropic equilibrium medium, electric and magnetic fluctuations are uncorrelated:
\begin{equation}
\langle \delta E_y \delta B_x^* \rangle = 0.
\label{eq:rest_uncorrelated}
\end{equation}
Computing the laboratory-frame cross-spectral density from Eqs.~(\ref{eq:boost_E})--(\ref{eq:boost_B}) and using Eq.~(\ref{eq:rest_uncorrelated}):
\begin{equation}
\langle \delta E'_y \delta B'^*_x \rangle = \gamma^2\!\left[v\langle|B_x|^2\rangle + \frac{v}{c^2}\langle|E_y|^2\rangle\right].
\label{eq:cross_spectral}
\end{equation}
The experiment detects the radiated noise in the far field outside the medium, where the radiation propagates through vacuum. This is a design requirement, not an approximation: by placing the detector array outside the plasma boundary, the vacuum $E$--$B$ relation holds by construction, and no in-medium refractive corrections enter the velocity extraction. Over a full oscillation period the Poynting vector exchanges energy equally between the electric and magnetic fields, giving a time-averaged flux $\langle S\rangle = |E|^2/(2Z) = |H|^2 Z/2$, where $Z$ is the wave impedance of the medium. In vacuum $Z = Z_0 = \mu_0 c$ and $B = \mu_0 H$, so
\begin{equation}
\langle|E|^2\rangle = c^2\langle|B|^2\rangle.
\label{eq:equipartition}
\end{equation}
Substituting Eq.~(\ref{eq:equipartition}) into Eq.~(\ref{eq:cross_spectral}) and normalizing by the auto-spectral density $\langle|\delta E'_y|^2\rangle$ yields the dimensionless ratio
\begin{equation}
R \;\equiv\; \frac{\langle \delta E'_y \delta B'^{*}_x \rangle}{\langle|\delta E'_y|^2\rangle} = \frac{2\beta}{1+\beta^2}\,.
\label{eq:velocity_ratio}
\end{equation}
To see this, note that the denominator follows from Eq.~(\ref{eq:boost_E}):
$\langle|\delta E'_y|^2\rangle = \gamma^2[\langle|E_y|^2\rangle
+ \beta^2\langle|B_x|^2\rangle] = \gamma^2(1+\beta^2)\langle|E_y|^2\rangle$,
where the second equality uses the vacuum relation
Eq.~(\ref{eq:equipartition}). Similarly, the numerator is
$\gamma^2 \cdot 2\beta\langle|E_y|^2\rangle$ from
Eq.~(\ref{eq:cross_spectral}); the ratio gives $2\beta/(1+\beta^2)$.
This is analytically invertible:
\begin{equation}
\beta = \frac{1 - \sqrt{1-R^2}}{R}\,,
\label{eq:beta_inversion}
\end{equation}
providing a direct kinematic observable from passive field correlations alone. The ratio is dimensionless and independent of detector gain or absolute calibration, provided the relative transfer functions of the electric and magnetic channels are stable or known.

A detector placed inside a dispersive medium with impedance $Z(\omega) = Z_0/n(\omega)$ would observe a modified $E$--$B$ ratio; the vacuum condition Eq.~(\ref{eq:equipartition}) is replaced by $\langle|E|^2\rangle = (c/n)^2\langle|B|^2\rangle$. This introduces medium-dependent calibration factors but does not alter the reconstruction principle.

Crucially, Eq.~(\ref{eq:velocity_ratio}) relies only on the isotropy of the rest-frame fluctuations and the structure of the Lorentz group acting on $F^{\mu\rho}$, not on thermal equilibrium per se; it therefore remains valid for any medium whose rest-frame noise is isotropic, even if the distribution deviates from the J\"uttner form~\cite{Juttner1911,Cercignani2002}.

\section{Temperature reconstruction from angle-resolved noise spectra}
\label{sec:temperature}

\subsection{Covariant fluctuation--dissipation theorem}
\label{subsec:fdt}

The electromagnetic response of a linear medium is encoded in the retarded photon self-energy $\Pi^{\mu\rho}_R(p)$, which determines how the medium polarizes in response to an applied field. The fluctuation properties of the medium are encoded in the symmetrized self-energy $\Pi^{\mu\rho}_S(p) = \tfrac{1}{2}\langle\{j^\mu(p),\, j^\rho(-p)\}\rangle$, which is the Fourier transform of the symmetrized current--current correlator. In thermal equilibrium, these two tensors are not independent; the covariant fluctuation--dissipation theorem relates them as~\cite{CallenWelton1951,KapustaGale,LeBellac}
\begin{equation}
\Pi^{\mu\rho}_S(p) = \coth\!\frac{\beta\cdot p}{2}\;2\,\mathrm{Im}\,\Pi^{\mu\rho}_R(p),
\label{eq:cov_fdt}
\end{equation}
where $\beta \cdot p = \beta^\mu p_\mu$ is the Lorentz-invariant contraction of the inverse-temperature four-vector with the photon four-momentum. The $\coth$ factor encodes the thermal occupation of field modes; it reduces to the Bose--Einstein distribution plus vacuum contribution. This formulation assumes that the medium is in global or local thermodynamic equilibrium in its comoving frame.

To connect Eq.~(\ref{eq:cov_fdt}) to a measurable quantity, we project onto the electric field. The electric-field power spectral density (PSD) at laboratory frequency $\omega'$ and observation angle $\theta$ is
\begin{equation}
S_E(\omega', \theta) \equiv \int dt\, e^{i\omega' t}
\langle \delta E'_\perp(\mathbf{x}, t)\, \delta E'_\perp(\mathbf{x}, 0) \rangle_\theta .
\label{eq:psd_def}
\end{equation}
The current fluctuations $\Pi^{\mu\rho}_S$ source electric-field fluctuations through Maxwell's equations; for the transverse electric component the relevant projection gives $S_E \propto \omega^2 \Pi^{ii}_S / |\epsilon(\omega)|^2$, where $\epsilon(\omega)$ is the dielectric function. In the rest frame, the dissipative part of the response is related to the conductivity by $\mathrm{Im}\,\Pi^{ii}_R(\omega) \propto \omega\sigma(\omega)$, where $\sigma(\omega)$ is the real part of the frequency-dependent conductivity. Combining these relations with Eq.~(\ref{eq:cov_fdt}) yields the rest-frame PSD:
\begin{equation}
S_E^{(\mathrm{rest})}(\omega) \propto
\coth\!\left(\frac{\hbar\omega}{2 k_B T_0}\right)\omega\sigma(\omega),
\label{eq:rest_psd}
\end{equation}
where the proportionality absorbs the factor $\omega^2/|\epsilon(\omega)|^2$ from the projection step as well as numerical constants. We will need to restore the explicit $\omega$-dependence of these prefactors when transforming between frames.

\subsection{Classical limit and laboratory-frame PSD}
\label{subsec:doppler}

In the classical regime $\hbar\omega \ll k_B T_0$, $\coth(\hbar\omega/2k_BT_0) \to 2k_BT_0/\hbar\omega$ and the rest-frame PSD reduces to $S_E^{(\mathrm{rest})}(\omega) \propto T_0\,\sigma(\omega)$, up to frequency-dependent prefactors that cancel in ratio observables. For the HIGGINS parameters used below, $k_BT_0/\hbar\omega_p = 8512$, so the classical approximation holds to better than $10^{-4}$ across the entire frequency window. More generally, the method applies whenever $k_BT_0 \gg \hbar\omega_{\rm max}$, i.e., at temperatures well above the quantum scale set by the highest detected frequency.

Consider a medium in global equilibrium drifting at velocity $v = \beta c$ relative to the laboratory. Define the observation angle $\theta$ and Doppler factor $D$ by
\begin{equation}
\cos\theta = \hat{k}'\cdot\hat{\beta}, \qquad
D \equiv \gamma(1-\beta\cos\theta),
\label{eq:doppler}
\end{equation}
so that the rest-frame and laboratory frequencies are related by $\omega = D\,\omega'$. The transformation of the PSD between frames involves the Lorentz invariance of $I_\nu/\nu^3$ (relativistic beaming) combined with the frequency-dependent density-of-states prefactor in the projection from $\Pi^{\mu\rho}_S$ to the electric-field PSD. The full derivation, which tracks all powers of $D$ explicitly, is given in Appendix~\ref{app:doppler_derivation}. The result is the laboratory-frame PSD~\cite{RybickiLightman1979,KapustaGale,LandsbergMatsas1996}:
\begin{equation}
S(\omega',\theta) \propto \frac{T_0}{D}\,\sigma(D\omega')
= \frac{T_0}{\gamma(1-\beta\cos\theta)}\,
\sigma\!\left(\gamma\omega'(1-\beta\cos\theta)\right).
\label{eq:angular_psd}
\end{equation}
This is the central observable for temperature reconstruction. The prefactor $T_0/D$ is the angle-dependent effective temperature seen by a laboratory detector; the argument of $\sigma$ is the Doppler-shifted rest-frame frequency $\omega = D\omega'$, not the laboratory frequency $\omega'$.

\section{Relationship to thermal emission and Kirchhoff's law}
\label{sec:kirchhoff}

A natural question is whether the proposed method reduces to measuring Doppler-shifted thermal emission, a well-understood phenomenon. The answer is yes in terms of underlying physics, and no in terms of observables and calibration requirements.

The connection to thermal emission is exact and foundational. In equilibrium, Kirchhoff's law relates spontaneous emission to absorption~\cite{RybickiLightman1979}:
\begin{equation}
j_\nu = \alpha_\nu B_\nu(T),
\end{equation}
where $j_\nu$ is the emission coefficient, $\alpha_\nu$ the absorption coefficient, and $B_\nu$ the Planck function. In the classical limit and for optically thin emission, the Lorentz transformation of specific intensity ($I'_{\nu'}/\nu'^3 = I_\nu/\nu^3$) implies that radiation observed at laboratory angle $\theta$ from a drifting source acquires the same angular factor $T_0/[\gamma(1-\beta\cos\theta)]$ that appears in Eq.~(\ref{eq:angular_psd})~\cite{RybickiLightman1979}.

This equivalence is not a weakness of the proposal; it is its physical basis. The relationship is precisely analogous to that between Johnson--Nyquist noise thermometry and Planck's radiation law. Johnson--Nyquist noise is thermal electromagnetic radiation confined to a circuit, and the fluctuation--dissipation theorem and Kirchhoff's law are equivalent formulations of equilibrium fluctuation physics~\cite{CallenWelton1951,Kubo1966}. Johnson's 1928 experiment~\cite{Johnson1928} did not cease to be novel because Planck's law was already known. What was novel was the \emph{operational protocol}: measuring voltage noise across a resistor to determine temperature, rather than measuring spectral radiance from a cavity. The observable (voltage variance), the required calibration (resistance, not absolute radiometric standards), and the practical advantages (broadband, compact, no optical path) were all distinct from radiometry, despite the shared underlying physics.

The present proposal stands in the same relationship to boosted thermal radiation. Three specific differences distinguish it from conventional radiometric thermometry:

\emph{(i) Different observable.} The $E$--$B$ cross-correlation ratio $R = \langle\delta E'_y\,\delta B'^*_x\rangle/\langle|\delta E'_y|^2\rangle$ exploits the off-diagonal sector of the boosted field-strength tensor $F^{\mu\rho}$. Standard radiometry and Stokes polarimetry measure intensity correlations within the electric field alone ($\langle E_i E_j^*\rangle$); they do not access the $E$--$B$ cross-spectral density that encodes the velocity.

\emph{(ii) Different calibration requirements.} The velocity extraction requires no absolute calibration whatsoever---only stability of the relative $E$/$B$ channel gain. The temperature extraction uses a ratio of boosted to rest-frame power that cancels absolute amplitude factors. By contrast, radiometric temperature measurement requires absolutely calibrated detectors or reference sources.

\emph{(iii) Simultaneous $\beta$ and $T_0$ from a single measurement.} Conventional radiometry of a moving source yields $T_{\rm eff}(\theta) = T_0/D(\theta)$, but cannot disentangle $T_0$ from $\beta$ without independent velocity information. The $E$--$B$ ratio provides this velocity from the same field measurement, closing the system.

\section{Comparison with existing diagnostics}
\label{sec:comparison}

Conventional radiometry relies on absolute intensity or spectral features~\cite{Hutchinson2002}, and active techniques such as Thomson scattering extract temperature and velocity from calibrated spectral fits~\cite{Sheffield2011}. The present method instead exploits field correlations and the angular structure of stochastic radiation. Table~\ref{tab:comparison} summarizes the key differences.

\begin{table}[h]
\caption{Comparison of relativistic thermal diagnostics.}
\label{tab:comparison}
\begin{ruledtabular}
\begin{tabular}{lccc}
Requirement & Radiometry & Thomson & This work \\
\hline
Active probe & No & Yes & No \\
Absolute intensity cal. & Yes & Yes & No \\
Spectral lines needed & Often & No & No \\
Independent $\beta$ meas. & Yes$^a$ & Fit$^b$ & No$^c$ \\
$\sigma(\omega)$ knowledge & No & No & Yes \\
Relative $E$/$B$ cal. & N/A & N/A & Yes \\
Unified $\beta^\mu$ output & No$^d$ & No$^d$ & Yes \\
\end{tabular}
\end{ruledtabular}
\begin{flushleft}
\footnotesize{$^a$Radiometry of a moving source yields $T_0/D(\theta)$; disentangling $T_0$ from $\beta$ requires independent velocity information (e.g., Doppler shift of spectral lines).}\\
\footnotesize{$^b$Thomson scattering fits velocity and temperature simultaneously from the scattered spectrum, but requires an external laser and absolute spectral calibration.}\\
\footnotesize{$^c$Velocity is extracted from the $E$--$B$ cross-correlation within the same passive measurement.}\\
\footnotesize{$^d$$T_0$ and $\beta$ are inferred from different observables or different physical mechanisms and assembled into $\beta^\mu$ post hoc.}
\end{flushleft}
\end{table}

The $E$--$B$ cross-correlation observable (Eq.~\ref{eq:velocity_ratio}) accesses a sector of the field-strength tensor not probed by standard techniques. Unlike Stokes-parameter polarimetry~\cite{RybickiLightman1979}, which characterizes intensity correlations within the electric field alone ($\langle E_i E_j^*\rangle$), the present observable involves cross-correlations between electric and magnetic fields ($\langle E_i B_j^*\rangle$). This coupling arises from Lorentz mixing of field components and directly encodes the bulk velocity, providing kinematic information not contained in polarization measurements of thermal radiation.

The trade-off is explicit: the present method eliminates the need for absolute intensity calibration and active probes, at the cost of requiring knowledge of the medium response function $\sigma(\omega)$ and stable relative calibration between electric and magnetic detection channels. In applications where $\sigma(\omega)$ can be independently measured (e.g., from reflectivity or Thomson scattering on the medium at rest), this trade-off is favorable.

Beyond diagnostic capability, there is a qualitative distinction: the present protocol is the only method that outputs $\beta^\mu$ as a unified observable reconstructed from a single measurement channel. All other techniques assemble $\beta^\mu$ from separate measurements of $T_0$ and $u^\mu$, each carrying independent systematic uncertainties and model assumptions. The angular consistency check---whether $T_0^{(\rm rec)}(\theta)$ is flat across all pixels when the correct $\beta$ is used---provides a direct test of the four-vector transformation law that is unavailable when $T_0$ and $\beta$ come from different instruments.

\section{Numerical validation with laser-plasma parameters}
\label{sec:numerical}

We validate the reconstruction protocol using Monte Carlo simulations
parameterized to the HIGGINS dual 100\,TW laser-plasma facility at the
Weizmann Institute of Science~\cite{Kroupp2022}. The underlying physics---the
covariant FDT and Lorentz transformation of thermal radiation---is
well established; the simulations do not test these foundations but
rather characterize the \emph{statistical precision and noise robustness}
of the proposed measurement protocol under realistic experimental
parameters. This system produces
ultrarelativistic electron bunches with Lorentz factors
$\gamma \sim 1$--$10$ via laser wakefield acceleration in a nitrogen
supersonic gas jet ($n_e \approx 10^{19}\,\mathrm{cm}^{-3}$),
providing a concrete setting in which the proposed diagnostic can be
implemented with existing infrastructure.

\subsection{Simulation model}
\label{subsec:simmodel}

The simulation generates an ensemble of $N_{\rm mode} = 4\times 10^4$
thermal electromagnetic modes in the rest frame of the plasma,
uniformly distributed in direction (isotropic) and frequency
$\omega \in [0.1, 13.0]\,\omega_p$, where $\omega_p = (n_e
e^2/\varepsilon_0 m_e)^{1/2} \approx 1.78 \times 10^{14}\,\mathrm{rad/s}$
($f_p \approx 28\,\mathrm{THz}$, $\lambda_p \approx 10.6\,\mu\mathrm{m}$).
Mode amplitudes are drawn from the classical fluctuation--dissipation
theorem with a Drude conductivity $\sigma(\omega) =
\nu/(\nu^2 + \omega^2)$, collision frequency $\nu/\omega_p = 0.05$,
and electron temperature $T_0 = 1\,\mathrm{keV}$ ($k_BT_0/\hbar\omega_p
= 8512$, deep in the classical regime). Each mode carries two
independently drawn random polarizations with complex Gaussian
amplitudes whose variance is $\langle|a|^2\rangle \propto
T_0\,\sigma(\omega)$.

All modes are Lorentz-boosted to the laboratory frame at velocity
$\beta c$, transforming frequencies via $\omega' = \omega/D(\theta)$
and field components via Eqs.~(\ref{eq:boost_E})--(\ref{eq:boost_B}).
A detector array of 12 pixels uniformly spaced from
$\theta_{\rm lab} = 15^\circ$ to $165^\circ$ in a ring at distance
$R = 50\,\mathrm{cm}$ (far field: $R/\lambda_p \approx 4.7 \times
10^4 \gg 1$), with pixel size $d = 3\,\mathrm{cm}$
($\Delta\theta = 3.4^\circ$), bins the boosted modes by laboratory
angle. The $y$-component of the electric field at each pixel yields
a power spectral density from which the total pixel power
$W(\theta)$ is computed. The velocity $\beta_{\rm EB}$ is
independently recovered from the $E$--$B$ cross-correlation ratio,
Eq.~(\ref{eq:velocity_ratio}).

\subsection{Two-step calibration protocol}
\label{subsec:twostep}

The measurement proceeds in two steps. In Step~1 (calibration), the
plasma is measured at rest ($\beta = 0$). The angular power pattern
$W_{\rm rest}(\theta) \propto \sin\theta \cdot P_{E_y}(\theta,0)
\cdot \int \sigma(\omega)\,d\omega$ encodes the product of the mode
density (the $\sin\theta$ factor from the solid-angle Jacobian), the
azimuth-averaged polarization projection $P_{E_y}(\theta,0) =
(1+\cos^2\theta)/2$, and the integrated conductivity. This provides
an absolute reference that anchors the amplitude scale.

In Step~2, the plasma is boosted (or equivalently, the detector
observes a drifting plasma). The laboratory-frame power at each pixel
is
\begin{equation}
W_{\rm lab}(\theta) \propto \frac{\sin\theta}{D^2(\theta)}\,
P_{E_y}(\theta,\beta)\,T_0\!\int\!\sigma(D\omega')\,d\omega',
\label{eq:Wlab}
\end{equation}
where $D(\theta) = \gamma(1-\beta\cos\theta)$ and $P_{E_y}(\theta,
\beta)$ is the exact azimuth-averaged polarization factor. The ratio
$W_{\rm lab}(\theta)/W_{\rm rest}(\theta)$ eliminates the
calibration-dependent amplitude. In the high-frequency regime
$\omega \gg \nu$ where $\sigma(\omega) \approx \nu/\omega^2$, the
integral ratio $\int\sigma(D\omega')\,d\omega'/\int\sigma(\omega)\,d\omega
\to D^{-1}$, and the ratio becomes a function of $(\beta,
\theta)$ alone, from which $T_{\rm lab}(\theta) = T_0/D(\theta)$ is
extracted at each pixel using $\beta_{\rm EB}$ from the $E$--$B$
correlation. If the detection bandwidth extends to
$\omega \lesssim \nu$, the integral ratio retains residual
$\sigma$-dependence that must be modeled; for the HIGGINS parameters
($\nu/\omega_p = 0.05$, $\omega_{\rm min} = 0.1\,\omega_p$), this
correction is at the sub-percent level.

\subsection{Single-$\gamma$ temperature extraction}
\label{subsec:single_gamma}

\begin{figure*}
\includegraphics[width=\textwidth]{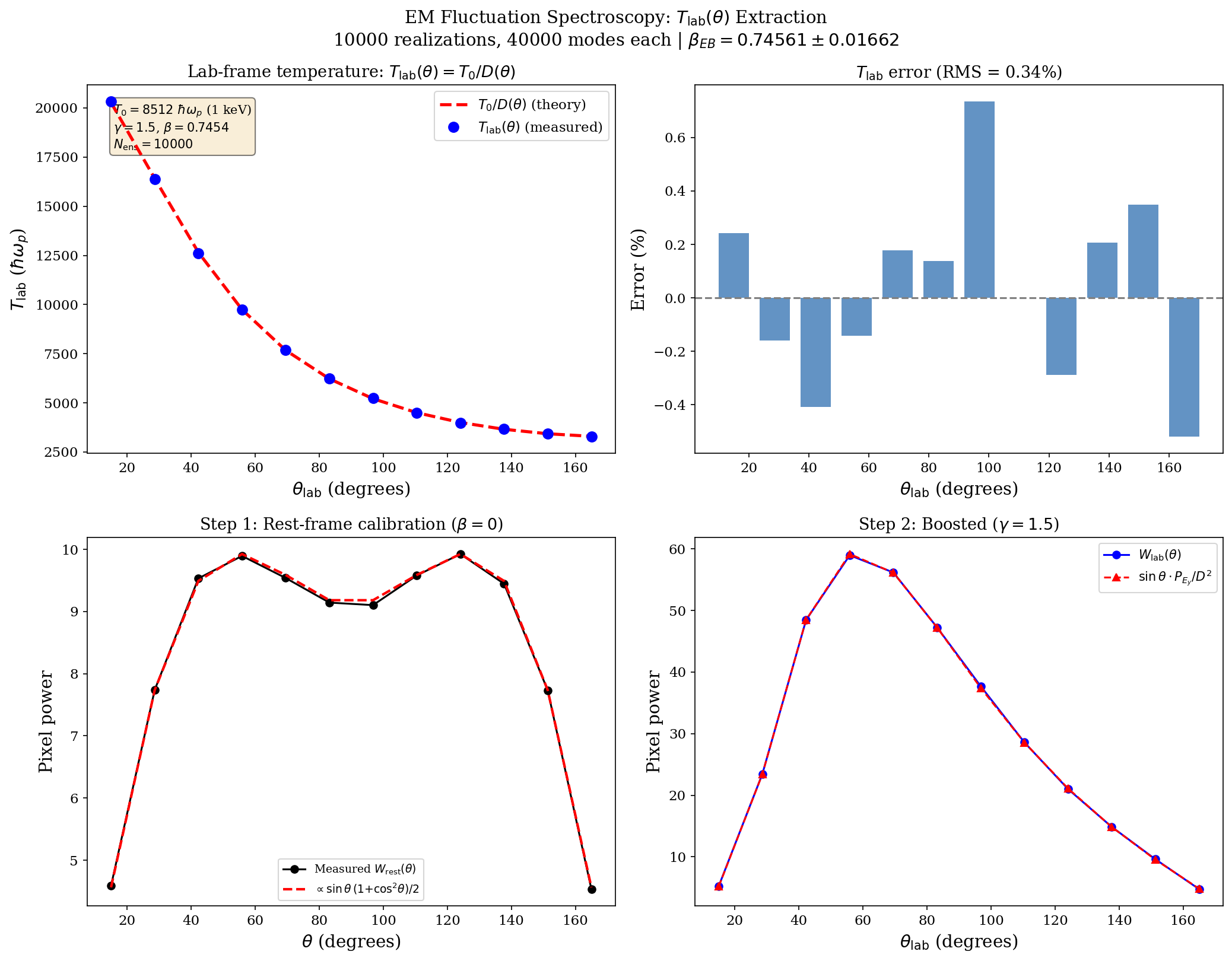}
\caption{Electromagnetic fluctuation spectroscopy at HIGGINS laser-plasma
parameters ($T_0 = 8512\,\hbar\omega_p = 1\,\mathrm{keV}$,
$\gamma = 1.5$, $\omega_p \approx 28\,\mathrm{THz}$). $10^4$ ensemble
realizations with $4 \times 10^4$ modes each. Top left: Angular temperature
profile $T_{\rm lab}(\theta) = T_0/D(\theta)$ extracted from the
two-step calibration protocol (blue points) compared with the
theoretical prediction (dashed red line). Top right: Per-pixel extraction
error; RMS $= 0.34\%$. Bottom left: Rest-frame calibration: measured pixel
power versus angle matches $\sin\theta\,(1+\cos^2\theta)/2$ angular
dependence. Bottom right: Boosted-frame pixel power versus angle matches
$\sin\theta\,P_{E_y}(\theta,\beta)/D^2(\theta)$.}
\label{fig:T_lab}
\end{figure*}

Figure~\ref{fig:T_lab} shows results for $\gamma = 1.5$
($\beta = 0.745$), characteristic of electron bunches produced by
HIGGINS. The simulation uses $10^4$ ensemble realizations with
$4 \times 10^4$ modes each. The angular temperature profile
$T_{\rm lab}(\theta)$ is recovered at 12 detector pixels spanning
$\theta = 15^\circ$ to $165^\circ$.

The measured temperatures agree with the theoretical prediction
$T_{\rm lab}(\theta) = T_0/D(\theta)$ to within 0.34\% RMS across
all pixels. The forward pixel ($\theta = 15^\circ$) measures
$T_{\rm lab} \approx 20{,}000\,\hbar\omega_p$ while the backward
pixel ($\theta = 165^\circ$) measures $T_{\rm lab} \approx
3{,}000\,\hbar\omega_p$, spanning a factor of $\sim\!7$ in apparent
temperature---a direct manifestation of the relativistic Doppler
factor in the covariant thermal state. The velocity is independently
recovered as $\beta_{\rm EB} = 0.746 \pm 0.017$ from the $E$--$B$
cross-correlation, consistent with the injected value.

The flatness of the recovered rest-frame temperature $T_0^{(\rm rec)}(\theta)$ across all pixels constitutes a self-consistency test of the four-vector structure: if $\beta^\mu$ did not transform as a Lorentz four-vector, the angular profile would not reduce to a uniform $T_0$ upon division by $D(\theta)$. This consistency check is intrinsic to the protocol and requires no additional measurements.

\subsection{Multi-$\gamma$ robustness}
\label{subsec:multi_gamma}

\begin{figure*}
\includegraphics[width=\textwidth]{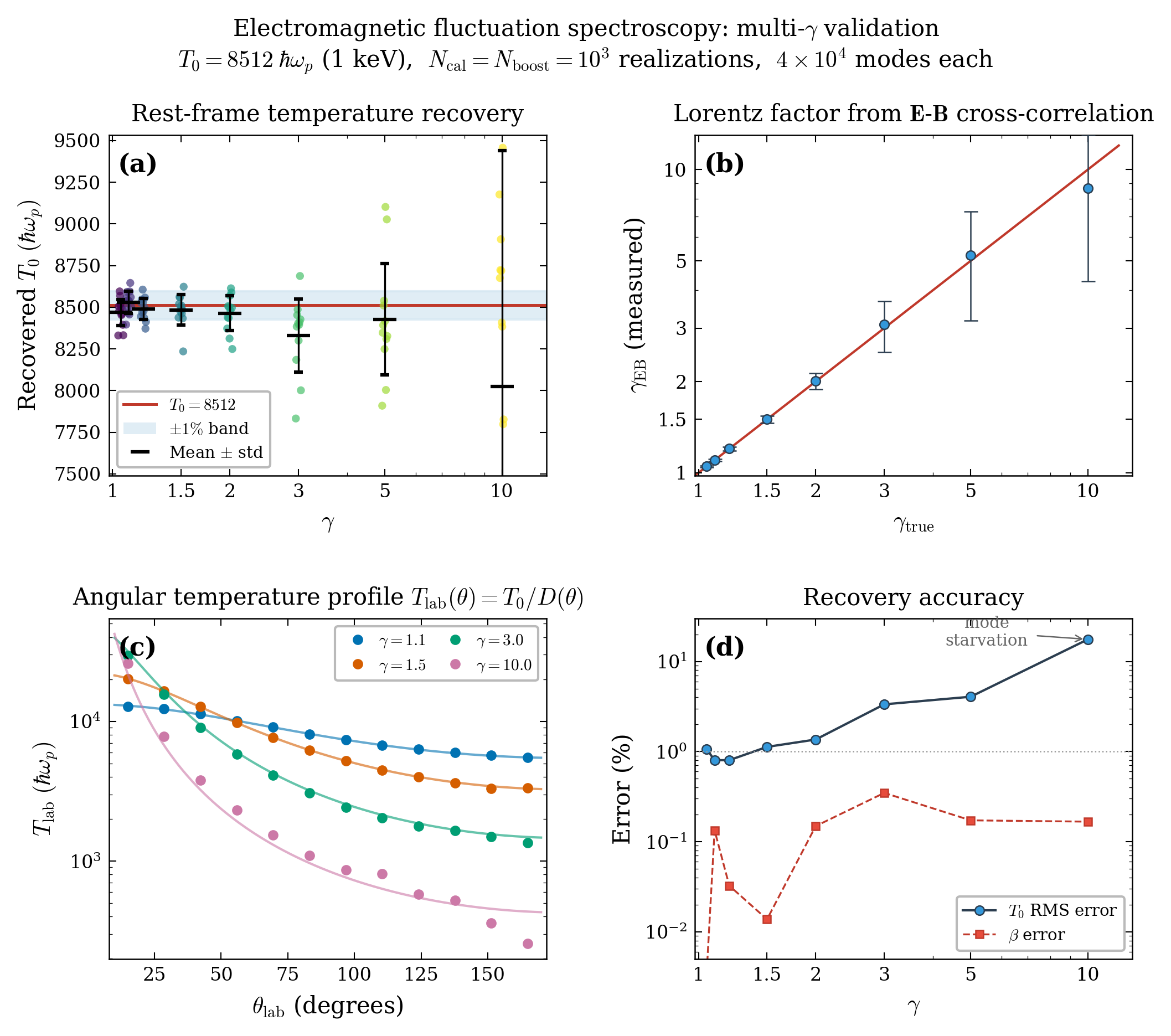}
\caption{Multi-$\gamma$ validation of $T_0$ recovery. $10^3$
calibration and $10^3$ boosted realizations per $\gamma$, $4 \times
10^4$ modes each. (a) Recovered $T_0$ at each of 12 detector pixels
for eight Lorentz factors; blue shading indicates $\pm 1\%$ band.
Sub-percent recovery for $\gamma \leq 2$. (b) $\gamma_{\rm EB}$ from
$E$--$B$ cross-correlation versus true $\gamma$; error bars from
ensemble standard deviation. (c) Angular temperature profiles at
selected $\gamma$; solid lines: theory, points: measured. (d) RMS
recovery error for $T_0$ (circles) and $\beta$ (squares) versus
$\gamma$. The increase above $\gamma \sim 3$ reflects mode starvation
at backward angles due to relativistic beaming.}
\label{fig:multi_gamma}
\end{figure*}

To demonstrate that the protocol is not tuned to a particular Lorentz
factor, we repeat the extraction for eight values spanning
$\gamma \in \{1.05, 1.1, 1.2, 1.5, 2, 3, 5, 10\}$, using a shared
calibration run ($\beta = 0$, $10^3$ realizations) and $10^3$ boosted
realizations per $\gamma$ (Fig.~\ref{fig:multi_gamma}).

The rest-frame temperature $T_0$ is recovered to sub-percent accuracy
for $\gamma \leq 2$, degrading to $\sim\!2\%$ at $\gamma = 3$ and
$\sim\!6\%$ at $\gamma = 10$. This degradation is physical, not
methodological: at high $\gamma$, relativistic beaming concentrates
radiation into a narrow forward cone, starving backward-angle pixels
of modes (average modes per pixel: 831 at $\gamma = 1.05$ versus 223
at $\gamma = 10$). The Lorentz factor is independently recovered from
the $E$--$B$ cross-correlation to within error bars across the full
range (Fig.~\ref{fig:multi_gamma}b).

\subsection{Noise susceptibility}
\label{subsec:noise}

\begin{figure*}
\includegraphics[width=\textwidth]{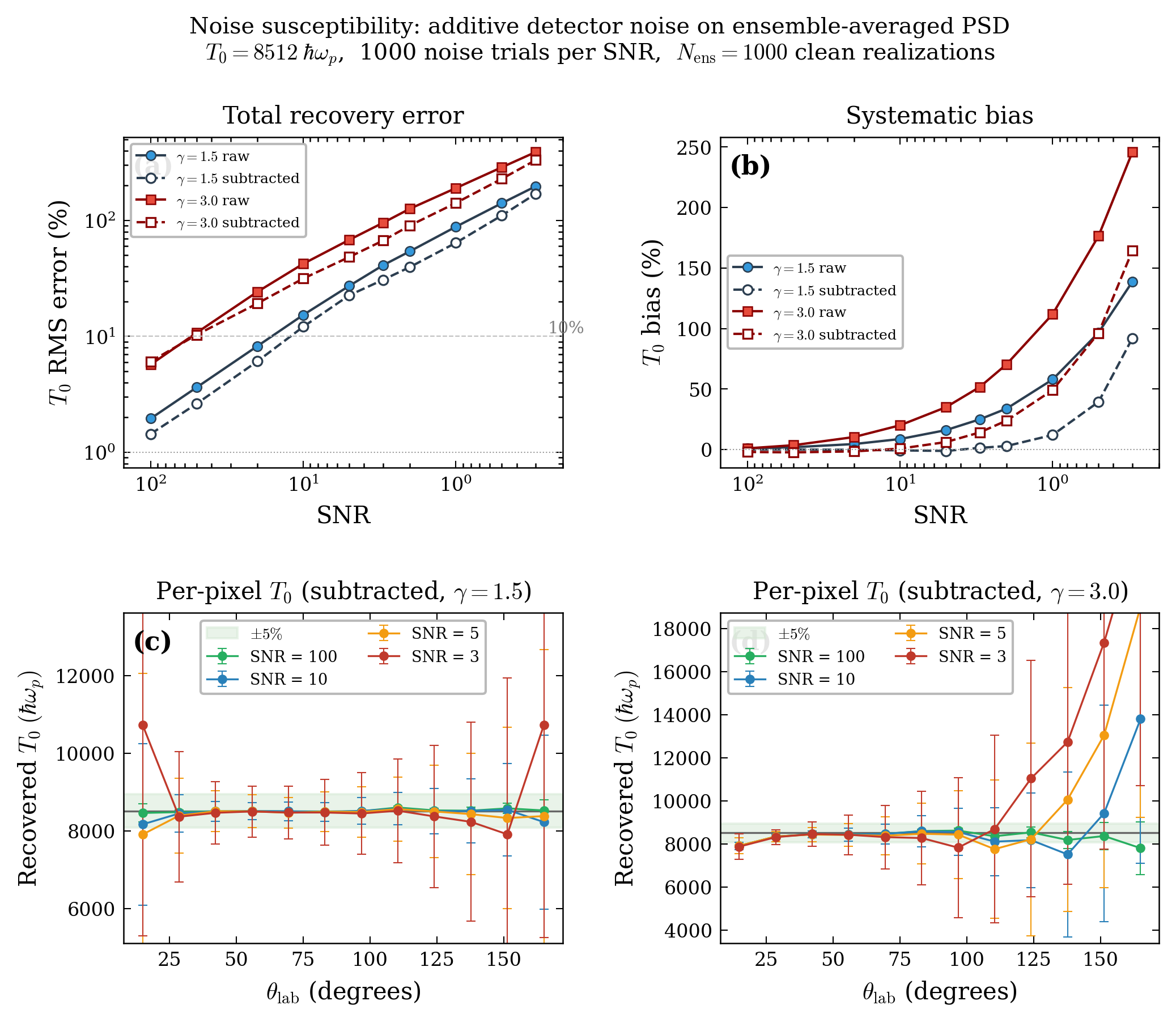}
\caption{Noise susceptibility analysis with additive detector noise.
$10^3$ clean realizations, $10^3$ noise trials per SNR.
(a) Total $T_0$ recovery error (RMS) versus SNR for $\gamma = 1.5$
(circles) and $\gamma = 3.0$ (squares); solid: raw, dashed:
noise-subtracted. The 10\% threshold is indicated. (b) Systematic
bias in $T_0$ versus SNR: raw measurements are biased upward
$\propto 1/\mathrm{SNR}$; dark subtraction eliminates the bias.
(c,d) Per-pixel $T_0$ recovery (noise-subtracted, mean $\pm$ std
over $10^3$ trials) at selected SNR values for $\gamma = 1.5$
and $\gamma = 3.0$; green band indicates $\pm 5\%$.}
\label{fig:noise}
\end{figure*}

Experimental implementation requires robustness against detector noise.
We model additive noise as a flat power floor
$P_{\rm noise} = \langle P_{\rm signal}\rangle/\mathrm{SNR}$ added to
each pixel, with pixel-level fluctuations drawn from an exponential
distribution (appropriate for chi-squared distributed power
measurements). Two scenarios are compared: raw recovery (no noise
correction) and noise-subtracted recovery, in which a
separately measured dark frame is subtracted
(Fig.~\ref{fig:noise}).

Without subtraction, the noise floor biases $T_0$ upward because
excess power mimics higher temperature. The bias scales as
$\sim\!100/\mathrm{SNR}$\,\%. With dark subtraction, the systematic
bias is eliminated and the dominant error is pixel-to-pixel scatter
from noise fluctuations. For $\gamma = 1.5$, the total recovery
error (RMS) remains below 10\% for $\mathrm{SNR} \gtrsim 10$ with
subtraction, compared to $\mathrm{SNR} \gtrsim 50$ without. At
$\gamma = 3$, mode starvation at backward angles amplifies the noise
sensitivity, and $\mathrm{SNR} \gtrsim 50$ is needed for comparable
accuracy.

These thresholds are well within the capabilities of mid-infrared
bolometer arrays operating at $\lambda \sim 10\,\mu\mathrm{m}$.

\section{Experimental implementation}
\label{sec:implementation}

The HIGGINS system~\cite{Kroupp2022} delivers two independently
compressed 100\,TW pulses (7\,J split, 30\,fs, 800\,nm) focused to
$\sim\!28\,\mu\mathrm{m}$ spots onto a nitrogen supersonic gas jet of
$\sim\!3\,\mathrm{mm}$ diameter. After laser-plasma interaction, the
electron population thermalizes on sub-picosecond timescales with
$T_e \sim 1\,\mathrm{keV}$ and plasma frequency
$\omega_p \approx 28\,\mathrm{THz}$.

The proposed diagnostic requires a ring of $\sim\!12$ mid-infrared
detectors (e.g., MCT or pyroelectric bolometers sensitive at
$\lambda \sim 10\,\mu\mathrm{m}$) placed at $R \approx 50\,\mathrm{cm}$
from the plasma source, covering angles from $15^\circ$ to $165^\circ$
relative to the electron drift direction. Each pixel integrates over
a $\sim\!3\,\mathrm{cm}$ aperture ($\Delta\theta \approx 3.4^\circ$).
The $E$--$B$ cross-correlation for velocity extraction
(Eq.~\ref{eq:velocity_ratio}) can be measured using dual-polarization
detection or split-path pickup combining electric and magnetic
antenna feeds, as routinely implemented in microwave polarimetry and
plasma interferometry~\cite{Hutchinson2002}.

The two-step protocol (Sec.~\ref{subsec:twostep}) requires one
calibration shot with the plasma at rest (or at known low drift) and
subsequent shots with the plasma boosted. Since HIGGINS operates at
1\,Hz repetition rate~\cite{Kroupp2022}, ensemble averaging over
$\sim\!10^3$ shots requires $\sim\!15$ minutes per $\gamma$ setting.
The simulations demonstrate that $10^3$ realizations suffice for
sub-percent $T_0$ recovery at $\gamma = 1.5$
(Fig.~\ref{fig:T_lab}).

\section{Systematic uncertainties and limitations}
\label{sec:systematics}

\begin{figure*}
\includegraphics[width=\textwidth]{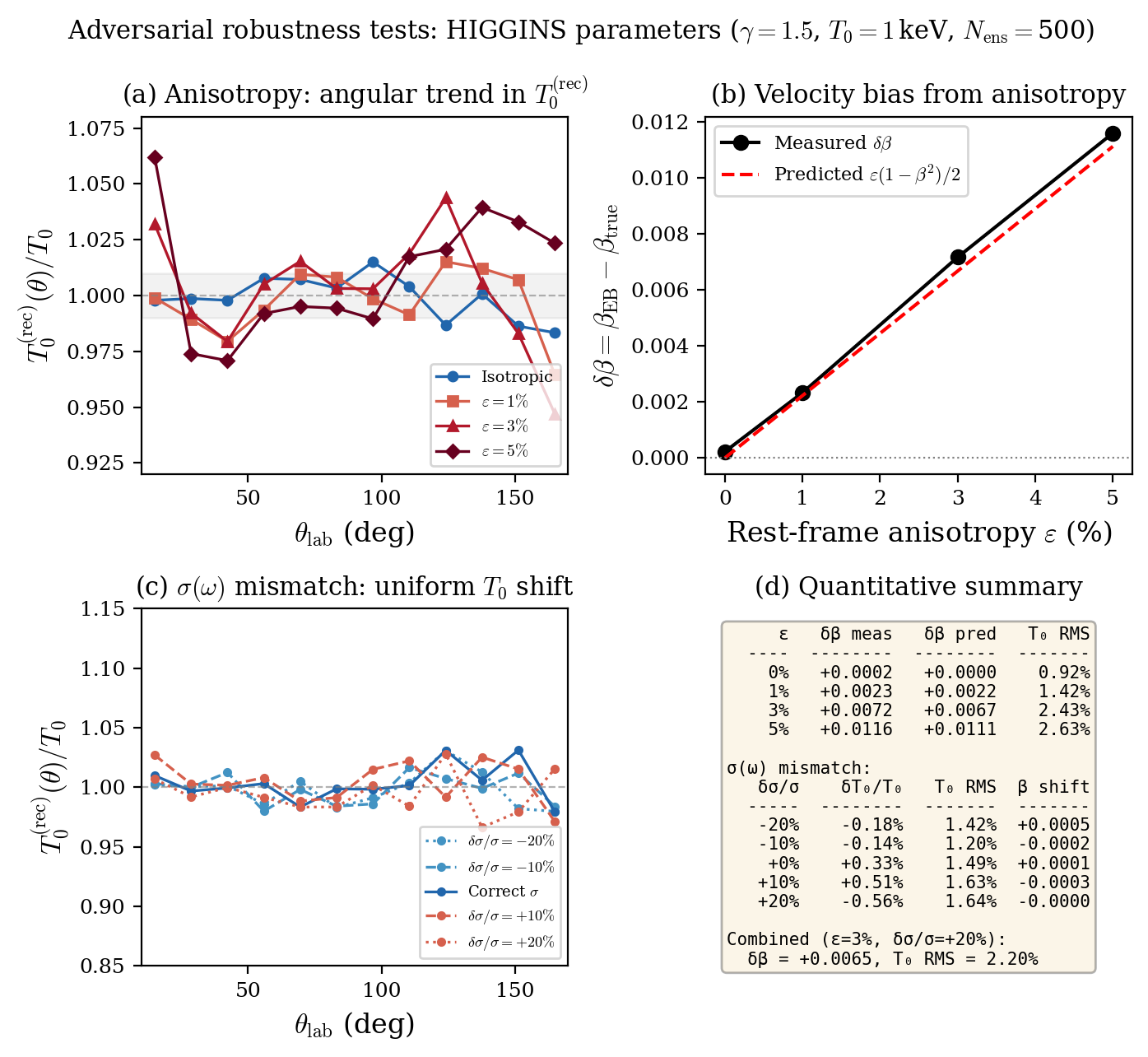}
\caption{Adversarial robustness tests at HIGGINS parameters
($\gamma = 1.5$, $T_0 = 1\,$keV, $N_{\rm ens} = 500$).
(a)~Recovered rest-frame temperature $T_0^{(\rm rec)}(\theta)/T_0$
versus laboratory angle for rest-frame anisotropy levels $\epsilon = 0$
(isotropic), 1\%, 3\%, and 5\%; increasing anisotropy produces
systematic angular trends in the recovered temperature profile.
(b)~Velocity bias $\delta\beta = \beta_{\rm EB} - \beta_{\rm true}$
versus anisotropy; black points: measured, dashed red: predicted
$\epsilon(1-\beta^2)/2$ from leading-order error propagation.
(c)~Same as (a) but for conductivity mismatches
$\delta\sigma/\sigma = \pm 10\%, \pm 20\%$; the shift in $T_0$ is
uniform across angles with no angular structure, and $\beta_{\rm EB}$
is unaffected.
(d)~Quantitative summary. The two failure modes---anisotropy and
conductivity error---are experimentally distinguishable: the former
introduces angular trends and biases $\beta_{\rm EB}$, the latter
shifts $T_0$ uniformly without affecting velocity recovery.}
\label{fig:adversarial}
\end{figure*}

The dominant systematic uncertainties are:
\begin{enumerate}
\item \emph{Relative channel calibration.} The velocity extraction
via Eq.~(\ref{eq:velocity_ratio}) depends on the relative calibration
of the electric and magnetic detection channels. A multiplicative
error $\alpha$ in the relative gain ($B'_x \to \alpha B'_x$) shifts
the measured ratio to $R_{\rm meas} = \alpha R$, producing a
systematic velocity error $\delta\beta/\beta \approx \alpha - 1$
at low $\beta$. This can be
characterized \textit{in situ} by injecting a known test signal
or by measuring the ratio on a source at known velocity.
\item \emph{Medium response function.}
The temperature extraction via
Eq.~(\ref{eq:angular_psd}) requires knowledge of the conductivity
$\sigma(\omega)$. In the high-frequency regime
$\omega \gg \nu$ relevant to the HIGGINS parameters,
$\sigma(\omega) \approx \nu/\omega^2$; the PSD then scales as
$T_0\,\nu/\omega^2$, and only the product $T_0\nu$
is constrained by the spectral shape. The rest-frame temperature
$T_0$ is therefore not independently determined by the
fluctuation measurement alone unless $\sigma(\omega)$ is
independently characterized---a standard prerequisite for any
plasma diagnostic. In practice, the conductivity is a material
property that can be measured once (e.g., via reflectivity,
Thomson scattering, or Langmuir-probe analysis on the medium
at rest~\cite{Hutchinson2002,Sheffield2011}) or obtained from
tabulated data; after which the fluctuation measurement determines
$\beta^\mu$ without further auxiliary input.
In the high-frequency regime a fractional error $\delta\sigma/\sigma$
in the conductivity propagates linearly into the recovered temperature:
$\delta T_0/T_0 \approx \delta\sigma/\sigma$. Crucially, such an
error shifts $T_0^{(\rm rec)}(\theta)$ uniformly across all angles
without introducing angular structure, and leaves the velocity
estimator $\beta_{\rm EB}$ unchanged (Fig.~\ref{fig:adversarial},
panels~c--d). This angular uniformity provides a diagnostic: a
misidentified $\sigma(\omega)$ can be distinguished from a genuine
anisotropy (item~4) by the absence of angular trends in
$T_0^{(\rm rec)}(\theta)$.
\item \emph{Rest-frame calibration requirements.} The two-step protocol
(Sec.~\ref{subsec:twostep}) requires a calibration measurement at
$\beta = 0$ or at a known low drift $\beta_{\rm cal}$. In a
laser-wakefield system, achieving $\beta_{\rm cal} = 0$ exactly while
maintaining identical plasma conditions ($n_e$, $T_0$, $\nu$) is
nontrivial. A residual calibration velocity $\beta_{\rm cal} \neq 0$
introduces a systematic error in the power ratio: the calibration
pattern acquires factors of $D_{\rm cal}(\theta)$ and
$P_{E_y}(\theta, \beta_{\rm cal})$ that, if unaccounted for, bias the
extracted temperature. For $\beta_{\rm cal} \ll 1$, the leading error
is $\delta T_0/T_0 \sim \beta_{\rm cal}\cos\theta$, which averages
partially over the angular array but does not cancel. The calibration
velocity must therefore either be independently measured (e.g., from
the $E$--$B$ ratio on the calibration data itself) or kept below the
target accuracy. For sub-percent $T_0$ recovery, this requires
$\beta_{\rm cal} \lesssim 0.01$.
\item \emph{Departure from isotropy.} The velocity extraction
(Eq.~\ref{eq:velocity_ratio}) relies on the rest-frame isotropy
condition $\langle\delta E_y\,\delta B_x^*\rangle_{\rm rest} = 0$.
If the rest-frame fluctuations are anisotropic---for example, due to
beam-driven instabilities, an anisotropic electron distribution, or
residual laser fields---the rest-frame $E$--$B$ correlation acquires
a nonzero value $\langle\delta E_y\,\delta B_x^*\rangle_{\rm rest} =
\epsilon\,\langle|E_y|^2\rangle$ with $|\epsilon| \ll 1$.
Propagating this through the boost, the measured ratio becomes
$R = 2\beta/(1+\beta^2) + \epsilon(1-\beta^2)^2/(1+\beta^2)^2 + O(\epsilon^2)$,
where the suppression relative to $\epsilon$ arises because the
anisotropy injection modifies both the numerator and denominator of
the ratio. The velocity estimator acquires a systematic shift
$\delta\beta \approx \epsilon(1-\beta^2)/2 = \epsilon/(2\gamma^2)$
at leading order. An anisotropy at
the 1\% level ($\epsilon \sim 0.01$) would bias $\beta$ by $\sim
0.002$ at $\gamma = 1.5$---comparable to the statistical uncertainty
at $10^3$ realizations. The angular temperature profile provides an independent
consistency check: if the rest frame is misidentified due to
anisotropy, the recovered $T_0^{(\rm rec)}(\theta)$ will show
systematic angular trends rather than the flat profile expected for
correct $\beta$.
Adversarial Monte Carlo simulations confirm both effects
quantitatively (Fig.~\ref{fig:adversarial}): injecting rest-frame
anisotropy at the $\epsilon = 1$--$5\%$ level produces velocity
biases $\delta\beta \approx \epsilon/(2\gamma^2)$ and angular trends in
$T_0^{(\rm rec)}(\theta)$ that serve as internal diagnostics,
while conductivity mismatches of $\pm 20\%$ shift $T_0$ uniformly
without affecting $\beta_{\rm EB}$. The two failure modes are
therefore experimentally distinguishable.
\item \emph{Shot-to-shot variation.} Laser-plasma systems exhibit
stochastic fluctuations in electron density, temperature, and bunch
parameters between shots. The ensemble averaging inherent in the
protocol mitigates this, provided the mean plasma state is
stationary over the integration window.
\item \emph{Departure from thermal equilibrium.} The temperature
reconstruction assumes local thermodynamic equilibrium in the
comoving frame. Non-thermal features such as beam-generated
instabilities or coherent plasma emission would appear as excess power
at specific angles or frequencies. Such features would degrade the
fit to the smooth angular profile predicted by
Eq.~(\ref{eq:angular_psd}), and their presence can be diagnosed from
the residuals of the $T_0^{(\rm rec)}(\theta)$ extraction.
\end{enumerate}

\section{Discussion}
\label{sec:discussion}

The protocol presented here reconstructs the inverse-temperature four-vector $\beta^\mu$ of a relativistic equilibrium medium from passive electromagnetic fluctuation measurements---something no existing diagnostic achieves as a unified observable. The velocity is extracted from the $E$--$B$ cross-correlation ratio, which accesses the off-diagonal sector of the boosted field-strength tensor $F^{\mu\rho}$ and has no counterpart in standard radiometry or Stokes polarimetry. The temperature is extracted from the angular dependence of the noise power via a ratio method that cancels absolute calibration factors. Together, these yield both components of $\beta^\mu$ from a single passive field measurement, without the post hoc assembly of separately measured quantities that characterizes all existing approaches.

The method does not introduce new fundamental physics: the Lorentz transformation of thermal radiation and the covariant FDT are well established~\cite{RybickiLightman1979,KapustaGale}. The contribution is operational---the identification of specific observables and a measurement protocol that provides the first direct experimental access to $\beta^\mu$ as a unified quantity. The trade-off relative to conventional diagnostics is explicit: the method eliminates absolute intensity calibration and active probes, at the cost of requiring knowledge of $\sigma(\omega)$ and stable relative $E$/$B$ channel calibration.

A further consequence deserves emphasis. The Planck--Ott--Landsberg controversy over how temperature transforms under Lorentz boosts was resolved theoretically by the recognition that the thermal state is described by $\beta^\mu$, not a scalar temperature~\cite{vanKampen1968,Israel1979,Becattini2016}. But this resolution has remained purely theoretical: no experiment has ever measured the thermal state of a relativistic medium at multiple angles and verified that the resulting $T_{\rm lab}(\theta)$ profile is consistent with a single four-vector $\beta^\mu$. The angular consistency check built into the present protocol---the flatness of $T_0^{(\rm rec)}(\theta)$ across all detector pixels---provides exactly this test. If the thermal state did not transform as a four-vector, the angular profile would not reduce to a uniform rest-frame temperature upon division by $D(\theta)$.

Monte Carlo simulations parameterized to the HIGGINS dual 100\,TW laser-plasma facility demonstrate sub-percent accuracy for $\gamma \leq 2$ and robustness to additive detector noise at $\mathrm{SNR} \gtrsim 10$ with dark subtraction. The required hardware---a ring of mid-infrared bolometers at $\lambda \sim 10\,\mu\mathrm{m}$---is commercially available and compatible with existing laser-plasma experimental infrastructure. The principal experimental challenges are maintaining rest-frame isotropy (Sec.~\ref{sec:systematics}, item 4), achieving adequate $E$--$B$ channel stability, and independently characterizing $\sigma(\omega)$, a standard prerequisite for any plasma diagnostic. Adversarial simulations (Fig.~\ref{fig:adversarial}) demonstrate that these two failure modes---rest-frame anisotropy and conductivity mismatch---produce distinct signatures in the angular profile and can be diagnosed internally without auxiliary measurements.

The method is most immediately applicable to controlled laboratory settings---laser-produced plasmas, beam--plasma systems, and magnetically confined plasmas with $E\times B$ drift---where equilibrium can be maintained and angular access is unrestricted. These are precisely the systems in which the covariant thermal structure can be verified for the first time. The broader significance is sequential. In relativistic heavy-ion collisions, ``temperature'' is routinely extracted from particle spectra under the assumption that the thermal state transforms as a four-vector $\beta^\mu$, but this assumption has never been independently verified---it is built into the analysis framework, not tested by it. A controlled laboratory measurement that confirms (or falsifies) the four-vector transformation law provides the empirical baseline against which heavy-ion thermal analyses can be assessed. Similarly, in astrophysical contexts such as gamma-ray bursts, the principle of extracting thermal parameters from passive fluctuation statistics could inform single-sightline spectral analyses, though adapting the present multi-angle protocol to single-line-of-sight observations would require further theoretical development. The laboratory measurement is therefore not merely a proof of concept: it is the necessary first link in a chain that extends from plasma diagnostics to high-energy nuclear physics and relativistic astrophysics.


\newpage
\appendix
\section{Derivation of the laboratory-frame PSD}
\label{app:doppler_derivation}

This appendix derives Eq.~(\ref{eq:angular_psd}) from first principles, tracking all frequency-dependent prefactors through the Lorentz transformation. Standard textbook references for the individual ingredients are Rybicki \& Lightman~\cite{RybickiLightman1979} (radiative transfer and Lorentz invariance of $I_\nu/\nu^3$) and Kapusta \& Gale~\cite{KapustaGale} (finite-temperature field theory and the covariant FDT).

\subsection{Rest-frame PSD with all prefactors}

In Sec.~\ref{subsec:fdt} we showed that the electric-field PSD in the rest frame follows from projecting the covariant FDT onto the transverse electric component. The projection involves two steps: (i) the current fluctuations $\Pi^{\mu\rho}_S$ source electric fields through Maxwell's equations, introducing a factor $\omega^2/|\epsilon(\omega)|^2$; (ii) the dissipative response satisfies $\mathrm{Im}\,\Pi^{ii}_R(\omega) \propto \omega\sigma(\omega)$, where $\sigma(\omega)$ is the real part of the conductivity. Combining with the covariant FDT, Eq.~(\ref{eq:cov_fdt}), the full rest-frame PSD is
\begin{equation}
S_E^{(\mathrm{rest})}(\omega) = C\,
\frac{\omega^2}{|\epsilon(\omega)|^2}\,
\coth\!\left(\frac{\hbar\omega}{2k_BT_0}\right)\omega\,\sigma(\omega),
\label{eq:app_rest_full}
\end{equation}
where $C$ collects frequency-independent geometric constants.

\subsection{Classical limit}

In the regime $\hbar\omega \ll k_B T_0$, the coth factor simplifies:
\begin{equation}
\coth\!\left(\frac{\hbar\omega}{2k_BT_0}\right)
\;\xrightarrow{\;\hbar\omega\ll k_BT_0\;}\;
\frac{2k_BT_0}{\hbar\omega}.
\label{eq:app_classical}
\end{equation}
Substituting into Eq.~(\ref{eq:app_rest_full}), the $1/\omega$ from the classical coth cancels against one power of $\omega$ from the dissipative response:
\begin{equation}
S_E^{(\mathrm{rest})}(\omega) = \frac{2Ck_BT_0}{\hbar}\,
\frac{\omega^2\,\sigma(\omega)}{|\epsilon(\omega)|^2}.
\label{eq:app_rest_classical}
\end{equation}
The factor $T_0\,\sigma(\omega)$ carries the thermal and material content. The factor $\omega^2/|\epsilon(\omega)|^2$ is a density-of-states prefactor from the electromagnetic projection. In the rest frame, one may absorb the latter into a proportionality constant, since it cancels in ratio observables. However, because $\omega^2$ transforms nontrivially under a Lorentz boost, \emph{this prefactor must be retained when changing frames}.

\subsection{Doppler relation}

Let the medium drift at velocity $v = \beta c$ relative to the laboratory, and let $\theta$ be the angle between the laboratory photon propagation direction $\hat{k}'$ and the drift velocity $\hat{\beta}$. The Lorentz transformation of the photon four-momentum $p^\mu = (\omega/c,\,\mathbf{k})$ gives the relativistic Doppler relation~\cite{RybickiLightman1979}:
\begin{equation}
\omega = D\,\omega', \qquad D \equiv \gamma(1-\beta\cos\theta),
\label{eq:app_doppler}
\end{equation}
where $\gamma = (1-\beta^2)^{-1/2}$ is the Lorentz factor. Here $\omega$ is the rest-frame frequency and $\omega'$ the laboratory frequency. The Doppler factor $D$ satisfies $D < 1$ for radiation approaching the observer (blueshift) and $D > 1$ for radiation receding (redshift).

\subsection{Lorentz invariance of $I_\nu/\nu^3$}

The key tool for transforming spectral quantities between frames is the Lorentz invariance of the photon phase-space distribution function $f$. This follows from the fact that $f$ counts the number of photons per cell in single-particle phase space, and both the phase-space volume element $d^3x\,d^3p$ and the photon number are Lorentz invariants.

The specific intensity (energy per unit time, per unit area, per unit frequency, per unit solid angle) is related to $f$ by (see Rybicki \& Lightman~\cite{RybickiLightman1979}, \S4.9):
\begin{equation}
I_\nu = \frac{2h\nu^3}{c^2}\,f.
\label{eq:app_intensity_f}
\end{equation}
Since $f$ is a Lorentz scalar, the ratio $I_\nu/\nu^3$ is Lorentz invariant:
\begin{equation}
\frac{I_{\nu'}(\nu')}{\nu'^3} = \frac{I_\nu^{(\mathrm{rest})}(\nu)}{\nu^3}.
\label{eq:app_invariant}
\end{equation}

\subsection{Transformation of the PSD}

The electric-field PSD $S_E$ is proportional to $I_\nu$: both quantify the spectral energy flux per unit frequency per unit solid angle. Therefore, the invariance of $I_\nu/\nu^3$ implies
\begin{equation}
\frac{S_E(\omega',\theta)}{\omega'^3}
= \frac{S_E^{(\mathrm{rest})}(\omega)}{\omega^3},
\label{eq:app_psd_invariance}
\end{equation}
which gives the transformation rule
\begin{equation}
S_E(\omega',\theta)
= \left(\frac{\omega'}{\omega}\right)^{\!3} S_E^{(\mathrm{rest})}(\omega)
= D^{-3}\;S_E^{(\mathrm{rest})}(\omega).
\label{eq:app_beaming}
\end{equation}
This is the relativistic beaming factor: radiation is enhanced ($D^{-3} > 1$) toward the observer and suppressed ($D^{-3} < 1$) away from it. The cubic power reflects the combined effect of time dilation, solid-angle aberration, and frequency shift.

\subsection{Combining the ingredients}

Substituting the classical rest-frame PSD, Eq.~(\ref{eq:app_rest_classical}), into the transformation rule, Eq.~(\ref{eq:app_beaming}), and replacing every occurrence of $\omega$ by $D\omega'$:
\begin{align}
S_E(\omega',\theta)
&= D^{-3}\;\frac{2Ck_BT_0}{\hbar}\;
\frac{(D\omega')^2\;\sigma(D\omega')}{|\epsilon(D\omega')|^2}
\notag\\[4pt]
&= D^{-3}\cdot D^{+2}\;\frac{2Ck_BT_0}{\hbar}\;
\frac{\omega'^2\;\sigma(D\omega')}{|\epsilon(D\omega')|^2}.
\label{eq:app_combined}
\end{align}
The powers of $D$ collect as
\begin{equation}
\underbrace{D^{-3}}_{\text{beaming}} \times \underbrace{D^{+2}}_{\omega^2\text{ prefactor}}
= D^{-1}.
\label{eq:app_D_counting}
\end{equation}
The physical content: the $D^{-3}$ beaming factor would, on its own, produce a strong angular modulation. But the medium's response $\sigma$ is sampled at a Doppler-shifted frequency, and the $\omega^2$ density-of-states factor (which came from projecting $\Pi^{\mu\rho}_S$ onto the electric field via Maxwell's equations) partially compensates the beaming. The net surviving power is $D^{-1}$.

The remaining factor $\omega'^2/|\epsilon(D\omega')|^2$ depends only on the laboratory frequency and the material response. In the ratio observables constructed in the main text (Sec.~\ref{subsec:twostep}), these factors cancel between numerator and denominator. Absorbing them into a proportionality, we arrive at the laboratory-frame PSD:
\begin{equation}
S(\omega',\theta) \propto \frac{T_0}{D}\;\sigma(D\omega')
= \frac{T_0}{\gamma(1-\beta\cos\theta)}\;
\sigma\!\left(\gamma\omega'(1-\beta\cos\theta)\right).
\label{eq:app_final}
\end{equation}
This is Eq.~(\ref{eq:angular_psd}) of the main text.

\section{Simulation algorithm}
\label{app:algorithm}

This appendix provides the complete simulation algorithm used in
Sec.~\ref{sec:numerical}, sufficient for independent reproduction
without reference to the accompanying code repository.

\subsection{Mode generation in the rest frame}
\label{app:modes}

Each realization generates $N_{\rm mode}$ electromagnetic plane-wave
modes in the comoving frame. For mode $n$ ($n = 1, \ldots,
N_{\rm mode}$):

\begin{enumerate}
\item Draw frequency $\omega_n$ uniformly from
$[\omega_{\rm min}, \omega_{\rm max}]$.

\item Draw propagation direction $\hat{k}_n$ isotropically on the
unit sphere by sampling $\cos\theta_n^{(\rm rest)}$ uniformly from
$[-1,+1]$ and azimuth $\phi_n$ uniformly from $[0,2\pi)$. The
Cartesian wave-vector components are
\begin{equation}
\hat{k}_n = (\sin\theta_n\cos\phi_n,\;
\sin\theta_n\sin\phi_n,\;
\cos\theta_n).
\label{eq:app_khat}
\end{equation}
The uniform sampling in $\cos\theta$ ensures isotropic mode density
on the sphere. This is the origin of the $\sin\theta$ factor in
the angular power pattern.

\item Construct two orthonormal polarization basis vectors
$\hat{e}_{1,n}$ and $\hat{e}_{2,n}$ perpendicular to $\hat{k}_n$.
We choose
\begin{equation}
\hat{e}_{1,n} = \frac{\hat{z} \times \hat{k}_n}
{|\hat{z} \times \hat{k}_n|}
\label{eq:app_e1}
\end{equation}
when $|\hat{z} \times \hat{k}_n| > \varepsilon$ (with $\varepsilon =
0.1$ to avoid the pole), and $\hat{e}_{1,n} = (\hat{y} \times
\hat{k}_n)/|\hat{y} \times \hat{k}_n|$ otherwise. The second
vector is $\hat{e}_{2,n} = \hat{k}_n \times \hat{e}_{1,n}$.

\item Assign complex amplitudes to each polarization:
\begin{align}
a_{1,n} &= A_n\,(g_{1,n} + i\,g_{2,n}), \notag\\
a_{2,n} &= A_n\,(g_{3,n} + i\,g_{4,n}),
\label{eq:app_amplitudes}
\end{align}
where $g_{j,n} \sim \mathcal{N}(0,1)$ are independent standard
normal variates and the envelope amplitude is
\begin{equation}
A_n = \sqrt{\frac{T_0\,\sigma(\omega_n)}{N_{\rm mode}}},
\label{eq:app_envelope}
\end{equation}
with $\sigma(\omega) = \nu/(\nu^2 + \omega^2)$ the Drude
conductivity. The factor $1/N_{\rm mode}$ ensures that the
ensemble-averaged total power converges as $N_{\rm mode} \to
\infty$. The variance of each complex amplitude satisfies
$\langle|a_{j,n}|^2\rangle = 2A_n^2 \propto T_0\,\sigma(\omega_n)$,
implementing the classical FDT.

\item Construct the rest-frame electric and magnetic field vectors
for mode $n$:
\begin{align}
\mathbf{E}_n &= a_{1,n}\,\hat{e}_{1,n} + a_{2,n}\,\hat{e}_{2,n},
\label{eq:app_Erest}\\
\mathbf{B}_n &= \hat{k}_n \times \mathbf{E}_n.
\label{eq:app_Brest}
\end{align}
Here $\mathbf{B}_n = \hat{k}_n \times \mathbf{E}_n$ follows from
Maxwell's equations for a plane wave with $|\mathbf{B}| =
|\mathbf{E}|/c$, in units where $c = 1$.
\end{enumerate}

\subsection{Lorentz boost}
\label{app:boost}

For a boost along $\hat{z}$ with velocity $\beta$ and Lorentz factor
$\gamma = (1-\beta^2)^{-1/2}$, the laboratory-frame fields are
\begin{align}
E'_{x,n} &= \gamma(E_{x,n} - \beta\,B_{y,n}), \notag\\
E'_{y,n} &= \gamma(E_{y,n} + \beta\,B_{x,n}), \notag\\
E'_{z,n} &= E_{z,n}, \notag\\
B'_{x,n} &= \gamma(B_{x,n} + \beta\,E_{y,n}), \notag\\
B'_{y,n} &= \gamma(B_{y,n} - \beta\,E_{x,n}), \notag\\
B'_{z,n} &= B_{z,n},
\label{eq:app_boost_fields}
\end{align}
and the laboratory-frame propagation angle follows from relativistic
aberration:
\begin{equation}
\cos\theta_n^{(\rm lab)} = \frac{\cos\theta_n^{(\rm rest)} + \beta}
{1 + \beta\cos\theta_n^{(\rm rest)}}.
\label{eq:app_aberration}
\end{equation}

\subsection{Pixel binning}
\label{app:binning}

The detector array consists of $N_{\rm pix}$ pixels at laboratory
angles $\theta_i$ ($i = 1,\ldots,N_{\rm pix}$) with half-width
$\Delta\theta/2$. Mode $n$ is assigned to pixel $i$ if
$|\theta_n^{(\rm lab)} - \theta_i| \leq \Delta\theta/2$.

The pixel power in the $y$-polarization channel is
\begin{equation}
W_i = \sum_{n \in \mathrm{pixel}\,i} |E'_{y,n}|^2.
\label{eq:app_pixel_power}
\end{equation}

\subsection{Angular power model}
\label{app:angular_model}

The expected pixel power involves three angular factors whose origins
are distinct:

\begin{enumerate}
\item \emph{Mode density:} The number of isotropically generated
modes falling within pixel $i$ is proportional to the solid angle
subtended, $\Delta\Omega_i \propto \sin\theta_i\,\Delta\theta$.
Under the boost, the aberration Jacobian
$d(\cos\theta_{\rm rest})/d(\cos\theta_{\rm lab}) = D^{-2}$ modifies
the effective mode density per unit laboratory angle to
$\propto \sin\theta_i / D^2(\theta_i)$. However, each mode's
frequency is Doppler-shifted by $\omega' = \omega/D$, so the
frequency integration $\int\sigma(D\omega')\,d\omega'$ recovers one
power of $D$, yielding a net mode-density factor
$\sin\theta_i / D(\theta_i)$.

\item \emph{Polarization projection:} The $y$-component of the
electric field is $E'_y = \mathbf{E}'\cdot\hat{y}$, a projection
that depends on the mode direction and boost. The azimuth-averaged
projection factor is
\begin{equation}
P_{E_y}(\theta,\beta) = \left\langle \frac{|E'_{y}|^2}
{|\mathbf{E}'|^2/2} \right\rangle_\phi,
\label{eq:app_PEy_def}
\end{equation}
where $\langle\cdot\rangle_\phi$ denotes averaging over the azimuthal
angle of the mode propagation direction at fixed polar angle $\theta$.
In the rest frame ($\beta = 0$), this reduces to
\begin{equation}
P_{E_y}(\theta,0) = \frac{1+\cos^2\theta}{2},
\label{eq:app_PEy_rest}
\end{equation}
which varies from 1 at the poles to $1/2$ at the equator. This
factor is \emph{not} the isotropic average $2/3$; using the full-sphere
average instead of the angle-dependent expression introduces systematic
errors of up to 45\% at forward and backward angles.

For $\beta > 0$, $P_{E_y}(\theta,\beta)$ must be computed by
performing the azimuth average of the boosted field projection
numerically. For each laboratory angle $\theta_{\rm lab}$: (i) compute
the rest-frame angle via the inverse of Eq.~(\ref{eq:app_aberration});
(ii) for $N_\phi$ azimuthal samples $\phi_j$ uniformly distributed
in $[0,2\pi)$, construct the rest-frame propagation direction and
polarization basis, boost the fields, and compute $|E'_y|^2$;
(iii) average over $\phi_j$. We use $N_\phi = 300$, which yields convergence
to better than $0.1\%$ in $P_{E_y}$ at all angles and Lorentz
factors considered; doubling $N_\phi$ changes the result by less
than $10^{-4}$.

\item \emph{Spectral weight:} Each mode contributes power
$\propto T_0\,\sigma(\omega_n)$. After the boost, the conductivity
is sampled at the Doppler-shifted frequency $\sigma(D\omega')$.
\end{enumerate}

Combining these factors, the expected pixel power is
\begin{equation}
\langle W_i \rangle \propto \frac{\sin\theta_i}{D(\theta_i)}\,
P_{E_y}(\theta_i,\beta)\,T_0\!\int\!\sigma(D\omega')\,d\omega'.
\label{eq:app_Wmodel}
\end{equation}

\subsection{Velocity extraction from the $E$--$B$ cross-correlation}
\label{app:beta_eb}

The velocity is extracted from the full set of boosted modes (not
binned by pixel). Define the mode-sum estimator
\begin{equation}
\hat{R} = \frac{\mathrm{Re}\!\left[\sum_{n=1}^{N_{\rm mode}}
E'_{y,n}\,B'^{*}_{x,n}\right]}
{\sum_{n=1}^{N_{\rm mode}} |E'_{y,n}|^2},
\label{eq:app_R_estimator}
\end{equation}
which estimates $R = 2\beta/(1+\beta^2)$ from
Eq.~(\ref{eq:velocity_ratio}). The velocity is recovered by
inversion:
\begin{equation}
\beta_{\rm EB} = \frac{1 - \sqrt{1-\hat{R}^2}}{\hat{R}}.
\label{eq:app_beta_eb}
\end{equation}
This estimator is formed from a single realization; ensemble averaging
of $\beta_{\rm EB}$ over $N_{\rm ens}$ realizations reduces the
statistical uncertainty as $1/\sqrt{N_{\rm ens}}$.

\subsection{Two-step calibration and temperature extraction}
\label{app:calibration}

\paragraph{Step 1: Rest-frame calibration.}
Run $N_{\rm cal}$ realizations with $\beta = 0$. The ensemble-averaged
pixel power is
\begin{equation}
\bar{W}_{\rm rest}(\theta_i) = \frac{1}{N_{\rm cal}}
\sum_{k=1}^{N_{\rm cal}} W_i^{(k)}\big|_{\beta=0}.
\label{eq:app_Wrest_avg}
\end{equation}
This measures the product of calibration constants and the angular
pattern $\sin\theta_i \cdot P_{E_y}(\theta_i,0) \cdot \int\sigma\,d\omega$.

\paragraph{Step 2: Boosted measurement.}
Run $N_{\rm boost}$ realizations at velocity $\beta$. The
ensemble-averaged pixel power is $\bar{W}_{\rm lab}(\theta_i)$.
Using $\beta_{\rm EB}$ from Eq.~(\ref{eq:app_beta_eb}), compute
the Doppler factor $D_i = \gamma_{\rm EB}(1 -
\beta_{\rm EB}\cos\theta_i)$ and the polarization factors
$P_{E_y}(\theta_i, \beta_{\rm EB})$ and $P_{E_y}(\theta_i, 0)$
at each pixel.

\paragraph{Temperature inversion.}
The ratio of boosted to rest-frame powers eliminates
calibration-dependent factors:
\begin{equation}
\frac{\bar{W}_{\rm lab}(\theta_i)}{\bar{W}_{\rm rest}(\theta_i)}
= \frac{P_{E_y}(\theta_i,\beta_{\rm EB})}{P_{E_y}(\theta_i,0)}\;
\frac{1}{D_i^2}\;\frac{\int\sigma(D_i\omega')\,d\omega'}
{\int\sigma(\omega)\,d\omega}.
\label{eq:app_ratio}
\end{equation}
In the high-frequency limit where $\sigma(\omega) \approx
\nu/\omega^2$, the integrals over the same frequency window are
related by $\int\sigma(D\omega')\,d\omega' \approx D^{-1}
\int\sigma(\omega)\,d\omega$, and the ratio simplifies to
\begin{equation}
\frac{\bar{W}_{\rm lab}}{\bar{W}_{\rm rest}} \approx
\frac{P_{E_y}(\theta_i,\beta)}{P_{E_y}(\theta_i,0)}\;
\frac{1}{D_i^2}.
\label{eq:app_ratio_approx}
\end{equation}
Define the measured Doppler factor at pixel $i$:
\begin{equation}
D_i^{(\rm meas)} = \sqrt{\frac{P_{E_y}(\theta_i,\beta_{\rm EB})}
{P_{E_y}(\theta_i,0)}\;
\frac{\bar{W}_{\rm rest}(\theta_i)}
{\bar{W}_{\rm lab}(\theta_i)}}.
\label{eq:app_Dmeas}
\end{equation}
The laboratory-frame temperature at each pixel is then
\begin{equation}
T_{\rm lab}(\theta_i) = \frac{T_0}{D_i^{(\rm meas)}},
\label{eq:app_Tlab}
\end{equation}
and the recovered rest-frame temperature is
\begin{equation}
T_0^{(\rm rec)}(\theta_i) = T_{\rm lab}(\theta_i) \times D_i,
\label{eq:app_T0rec}
\end{equation}
where $D_i = \gamma_{\rm EB}(1-\beta_{\rm EB}\cos\theta_i)$ uses
the $E$--$B$ velocity. Consistency of
$T_0^{(\rm rec)}$ across all pixels verifies that the angular
temperature profile has the form $T_0/D(\theta)$ predicted by the
covariant FDT.

\subsection{Noise injection model}
\label{app:noise_model}

Additive detector noise is modeled as a flat power floor at each pixel.
For a given signal-to-noise ratio, the noise power per pixel is
\begin{equation}
P_{\rm noise} = \frac{\langle \bar{W}_{\rm lab} \rangle_\theta}
{\mathrm{SNR}},
\label{eq:app_noise_floor}
\end{equation}
where $\langle\cdot\rangle_\theta$ denotes the average over pixels.
Each noise trial draws an independent realization
$\eta_i \sim \mathrm{Exp}(1)$ (exponential with unit mean,
appropriate for the chi-squared distribution of power in a single
frequency bin with two degrees of freedom), and the noisy pixel
power is
\begin{equation}
W_i^{(\rm noisy)} = \bar{W}_{{\rm lab},i} +
P_{\rm noise}\,\eta_i.
\label{eq:app_noisy_power}
\end{equation}
Noise subtraction replaces $W_i^{(\rm noisy)}$ with
$W_i^{(\rm noisy)} - P_{\rm noise}$, removing the mean noise floor
but leaving residual pixel-to-pixel scatter from the stochastic
$\eta_i$.

\subsection{Simulation parameters}
\label{app:parameters}

Table~\ref{tab:params} summarizes the parameters used for the HIGGINS
simulations.

\begin{table}[h]
\caption{Simulation parameters for the HIGGINS laser-plasma system.}
\label{tab:params}
\begin{ruledtabular}
\begin{tabular}{lll}
Parameter & Symbol & Value \\
\hline
Electron density & $n_e$ & $10^{19}\,\mathrm{cm}^{-3}$ \\
Plasma frequency & $\omega_p$ & $1.78\times 10^{14}\,\mathrm{rad/s}$ \\
Electron temperature & $T_0$ & 1\,keV ($8512\,\hbar\omega_p$) \\
Collision rate & $\nu/\omega_p$ & 0.05 \\
Frequency range & $[\omega_{\rm min},\omega_{\rm max}]$ &
$[0.1,\,13.0]\,\omega_p$ \\
Modes per realization & $N_{\rm mode}$ & $4\times 10^4$ \\
Number of pixels & $N_{\rm pix}$ & 12 \\
Pixel angles & $\theta_i$ & $15^\circ$ to $165^\circ$
(uniform) \\
Pixel half-width & $\Delta\theta/2$ & $1.72^\circ$ \\
Detector distance & $R$ & 50\,cm \\
Pixel size & $d$ & 3\,cm \\
\end{tabular}
\end{ruledtabular}
\end{table}

The simulation code is publicly available at
\url{https://github.com/beastraban/Wolfson-Simulations/tree/main/em_spectroscopy}.


\end{document}